\documentclass[10pt,twocolumn,superscriptaddress,aps,pra,balancelastpage]{revtex4-1}
\usepackage[utf8]{inputenc}
\usepackage{amsmath,amsfonts,amssymb,fullpage,color,graphicx,xcolor,physics}
\usepackage[colorlinks]{hyperref}

\begin{document}
\title{Distinguishing between quantum and classical Markovian dephasing dissipation}

\author{Alireza Seif}
\affiliation{Pritzker School of Molecular Engineering, University of Chicago, Chicago, IL 60637}
\author{Yu-Xin Wang}
\affiliation{Pritzker School of Molecular Engineering, University of Chicago, Chicago, IL 60637}
\author{Aashish A. Clerk}
\affiliation{Pritzker School of Molecular Engineering, University of Chicago, Chicago, IL 60637}
\date{April 2021}
\begin{abstract}
    Understanding whether dissipation in an open quantum system is truly quantum is a question of both fundamental and practical interest.  
    We consider $n$ qubits subject to correlated Markovian dephasing, and present a sufficient condition for when bath-induced dissipation can generate system entanglement and hence must be considered quantum.  Surprisingly, we find that the presence or absence of time-reversal symmetry (TRS) plays a crucial role: 
    broken TRS is required for dissipative entanglement generation.  
    Further, simply having non-zero bath susceptibilities is not enough for the dissipation to be quantum.  Our work also presents an explicit experimental protocol for identifying truly quantum dephasing dissipation, and lays the groundwork for studying more complex dissipative systems and finding optimal noise mitigating strategies.   
\end{abstract}
\maketitle

{\it Introduction-- } Open quantum systems, where a system of interest interacts with an external environment, play a central role in many areas of physics ranging from quantum information to cosmology.  Their evolution is in general non-unitary~\cite{breuer2002theory}, being described by e.g.~a quantum master equation or more general quantum map.  A ubiquitous type of system-environment interaction is dephasing. Dephasing interactions do not change populations of energy eigenstates of the system, but rather only impact coherences, i.e., the off-diagonal elements of the density matrix in the energy eigenbasis. The role of dephasing in quantum computation has been extensively studied~(see e.g.~\cite{palma1996quantum,duan1998reducing,reina2002decoherence}).  Even in this simple setting, there is a fundamental, surprisingly subtle question of interest: can the environment-induced dissipation of the system be attributed to interaction with a completely classical environment, or does it necessarily require a truly quantum environment? 

Answering this question is of course contingent on how one defines the line between classical and quantum environments. Several previous works have examined this issue (see e.g.~\cite{nori2018,nori2019,franco2019,viola2019,clerk2020,cywinski2020,chen2021}), 
largely in terms of possibly representing dissipative quantum dynamics with an equivalent classical process. In this work, we take instead an operational and experimentally-motivated approach, and define a truly quantum environment to be one that can mediate dissipative interactions that generate system entanglement. A necessary requirement for this phenomenon is having non-zero bath response susceptibilities~\cite{clerk2010qunoise,paz2017multiqubit,von2020two}, manifested in asymmetric-in-frequency environmental quantum noise spectral densities.
We show, surprisingly, that this is not sufficient:  bath-mediated dissipative interactions can only generate system entanglement if they cannot be mimicked using local measurements and feedforward.
We note that an analogous approach has been suggested to test probe whether graviational interactions are quantum~\cite{kafri2014classical,carney2021using}.

\begin{figure}[t]
    \centering
    \includegraphics[width=\columnwidth]{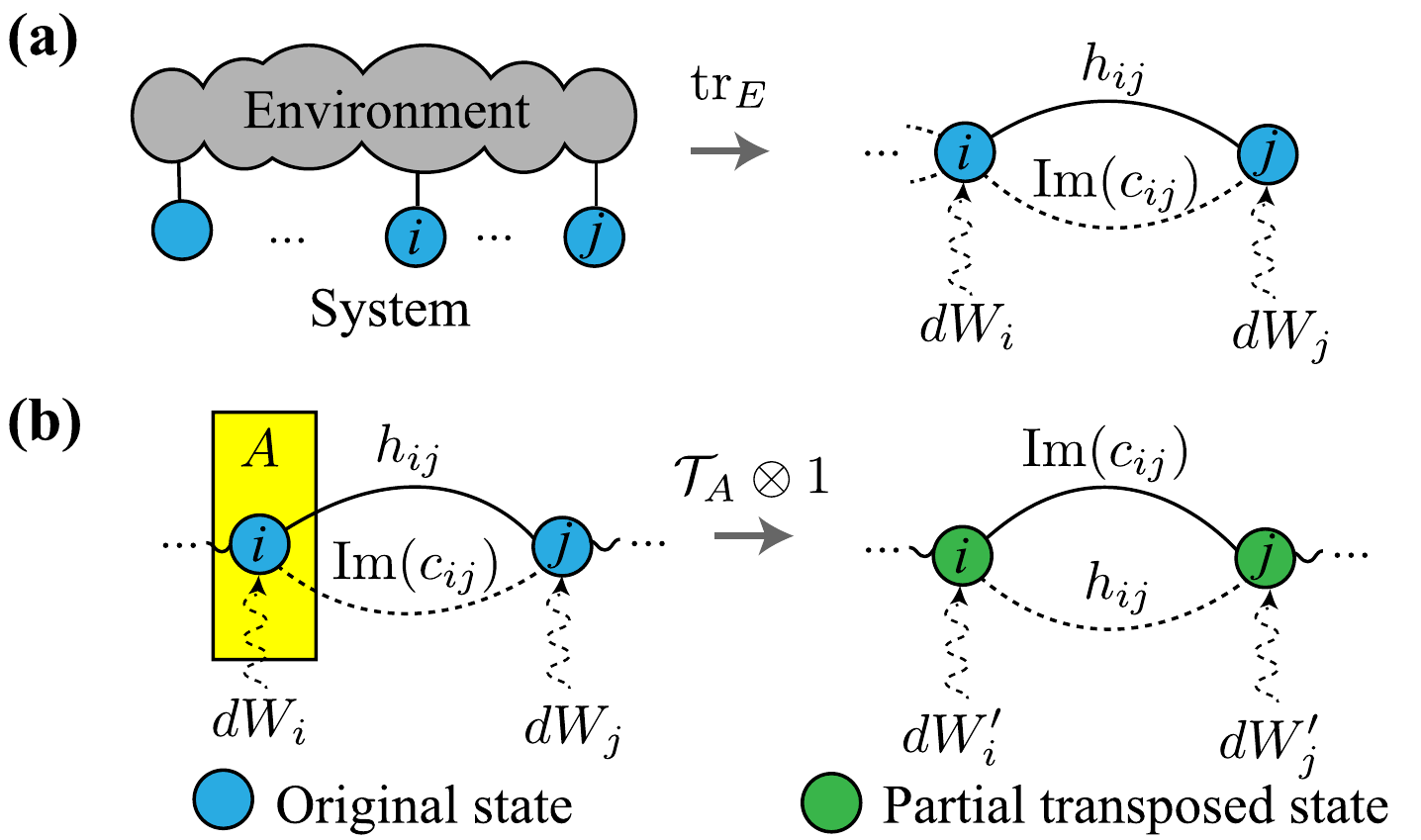}
    \caption{
    (a)  A system of $n$ qubits is coupled to a Markovian dephasing environment.  The resulting evolution obtained by tracing out the environment
    ($\tr_E$) 
    is a GKSL master equation (Eq.~\eqref{eq:lindblad_full}), and can be decomposed into driving by effective classical noise $dW_i$ with correlations $\Re(c_{ij})$, Hamiltonian Ising interactions $h_{ij}$, and dissipative Ising interactions $\Im(c_{ij})$ (b) The evolution of the partial transposed ($\mathcal{T}_A\otimes1$) state with respect to subsystem $A$ is again in GKSL form, with different classical noise $dW'_i$ whose correlation is now $-c_{ij}$ and with the role of $h_{ij}$ and $\Im(c_{ij})$ reversed.}
    \label{fig:schematic}
\end{figure}

We focus in this work on setups where a set of qubits are coupled to a generalized Markovian dephasing environment (as described by a Lindblad master equation).
We show that the presence or absence of environmental time-reversal symmetry (TRS) is crucial in determining whether bath-induced dissipation is classical.  In the presence of TRS, this dissipation is always equivalent to driving by classical noise, whereas without TRS this is no longer necessarily true.  We provide a condition based on Peres-Horodecki criterion~\cite{peres1996separability,horodecki2001separability} that allows one to identify truly quantum dephasing dissipation.  Our work provides a new approach to identifying truly quantum dissipative behaviour, and also provides a sensitive method for detecting broken TRS in dephasing environments.


{\it Setup-- }
Consider a multi-qubit system, whose dephasing interaction with a stationary environment is  $\hat{H}_{\rm{int}}=\sum_i \hat{Z}_i\otimes\hat{B}_i$, where $\hat{Z}_i$ is the Pauli $\hat{\sigma}_z = \ket{0} \bra{0} - \ket{1} \bra{1} $ operator on qubit $i$, and $\hat{B}_i$ is a Hermitian environment operator. Throughout the paper, we transform to the intraction picture with respect to the internal Hamiltonians of the system (S) and environment (E),  $\hat{H}_{\rm{S}}+\hat{H}_{\rm{E}}$, where the environment operators are given by $ \hat{B}_i(t) = e^{i \hat{H}_{\rm{E}} t } \hat{B}_i e^{-i \hat{H}_{\rm{E}} t }$.  In the Markovian limit, the evolution of the system undergoing such correlated dephasing is described by the  Gorini-Kossakowski-Sudarshan-Lindblad (GKSL) equation $\frac{d{\hat{\rho}}}{d t}= \mathcal{L}(\hat{\rho})$, with the Liouvillian given by
\begin{equation}\label{eq:lindblad_full}
    \mathcal{L}(\hat{\rho})= -i[\hat{H}_{\rm{LS}},\hat{\rho}] + \sum_{i,j} c_{ij} 
    \left(\hat{Z}_i \hat{\rho} \hat{Z}_j - \frac{1}{2} \{\hat{Z}_i \hat{Z}_j, \hat{\rho}\} \right).
\end{equation}
Here $\hat{H}_{\rm{LS}}=\frac{1}{2}\sum_{ij} h_{ij} \hat{Z}_i \hat{Z}_j$ is the so-called Lamb shift Hamiltonian (see Fig.~\ref{fig:schematic}a), and describes Hamiltonian Ising interactions mediated by the bath.  The remaining terms describe bath-induced system dissipation, and cause the net evolution to be non-unitary.
The evolution generated by this equation is completely positive (CP) if and only if the matrix $C=(c_{ij})$ is positive semi-definite (PSD)~\cite{lindblad1976generators}.

Our central goal is to understand whether the bath-induced dissipation is classical or quantum; we thus set $h_{ij}=0$ in what follows. 
It is tempting to think of the remaining dissipative evolution as always being equivalent to driving by external classical noise.  This is not true however if $\textrm{Im } C \neq 0$:  as we show below, the imaginary part of $C$ encodes dissipative bath-mediated interactions that are distinct from classical noise, and cannot be mimicked by a Hamiltonian Ising interaction.  Further, the presence of these interactions is necessary {\it but not sufficient} to make the dissipation quantum, i.e.~capable of generating system entanglement.

Consider first  Eq.~\eqref{eq:lindblad_full} with a real $C$ and $\hat{H}_{\rm{LS}}=0$.  Such an evolution can always be emulated by (correlated) classical white noise~\cite{van1992stochastic}. Specifically, consider a system evolving under the Hamiltonian $\hat{H}_c=\sum_i b_i(t) \hat{Z}_i$, where $b_i(t)$ describe classical Gaussian fluctuations with $\expval{b_i(t)}=0$. In the white noise limit, where $\expval{b_i(t)b_j(t)}=c_{ij} \delta(t-t')$, the average evolution of the system (over the fluctuations) reproduces the master equation of interest, with $C$ corresponding to the covariance matrix of noise $b_i(t)$~\cite{gardiner2004quantum,supplement}. Note that because this evolution can always be emulated by a local time-dependent Hamiltonian, it cannot create entanglement in the system. 

{\it Interpretation of complex $C$-- }
As shown above, driving a system with classical noise will never generate a non-zero $\Im(C)$. To understand the physical origin of $\Im(C)\neq 0$, we revisit the general microscopic quantum bath model and $\hat{H}_{\rm{int}}$. Making use of the standard Born-Markov approximation~\cite{breuer2002theory}, we can relate $\text{Im}(C)$ to spatial asymmetries in the environment's response properties: 
$ \text{Im}( 
c _{ jk}) =\frac{1}{2}
( \text{Re}[\chi _{ jk}  (\omega =0)]  
-  \text{Re}[\chi _{ kj }  (\omega =0)] ) $. Here,  
$\chi _{ jk}  (\omega) \equiv 
 - i  \int ^{\infty}_{ 0 } d t  
e^{i \omega t } \langle   [ 
\hat B  _j  (t) 
,\hat B  _k (0) 
]  \rangle $ are standard linear response susceptibilities, which describe how a bath operator expectation value $ \langle  \hat B _j \left ( t \right ) \rangle $ changes due to a weak perturbation that couples to $ \hat B _k$ ~\cite{clerk2010qunoise}.
In contrast, $\Re(C)$ is not related to bath response; for a generic quantum bath, $\Re(C)$ are given by the symmetrized quantum noise spectra, which play the role of classical noise~\cite{clerk2010qunoise}.
It is important to note that if the bath Hamiltonian $ \hat H _\mathrm{E} $ has time reversal symmetry (TRS) and the bath operators all transform under TRS with the same parity, we may invoke Onsager-type recriprocity relations to show that $ \text{Im}( 
c _{ jk} ) = 0$~\cite{onsager1931}. Conversely, for environments with broken TRS, which is typically the case for driven-dissipative environments, there is no fundamental reason to expect that $\text{Im}(C)$ should vanish.  

An alternate way to understand the bath-induced dissipation and $ \text{Im}( C ) $ is to realize that it also describes a situation where there is no environment, but where the system evolves because of continuous measurement and feed-forward \cite{wiseman_milburn_2009}.  We first write the dissipative part of Eq.~\eqref{eq:lindblad_full} as
\begin{equation}\label{eq:lindblad_diss}
    \mathcal{L}_{\rm{diss}}(\hat{\rho}) =
        \sum_k \gamma_k \left(
            \hat{L}_k \hat{\rho} \hat{L}_k^\dagger -\frac{1}{2} \{\hat{L}_k^\dagger \hat{L}_k,\hat{\rho}\}
            \right),
\end{equation}
where $\gamma_k$ and $\hat{L}_k$ are found by diagonalizing $C$~\cite{breuer2002theory}. Further, each $\hat{L}_k$ can be written as $\hat{L}_k=\hat{A}_k+i\hat{B}_k$ where $\hat{A}_k$,   $\hat{B}_k$ are Hermitian system operators.  While such dissipators arise in many contexts~\cite{gardiner1993driving,carmichael1993quantum,wiseman_milburn_2009,gough2009series}, we focus on the connection to continuous measurement and feedforward of Ref.~\cite{metelmann2017nonreciprocal}  As shown in the Supplementary Material (SM)~\cite{supplement}, each term $k$ in Eq.~\eqref{eq:lindblad_diss} describes unconditional system evolution under a two-way measurement and feedforward scheme.  One half of this scheme is a weak continuous measurement of $\hat A_k$~\cite{wiseman_milburn_2009}, with the measurement record used to set the amplitude of a drive applied to $-\hat B_k$.  The other half is the reverse process, i.e.~measuring $\hat B_k$ and feed-forwarding the result to drive $\hat A_k$.  

This equivalence immediately lets us make general statements on the properties of the dissipative evolution.  In general, the measurement and feed-forward realization involves nonlocal operations, indicating that a complex $C$ might create system entanglement. However, there are notable exceptions. First, emulating a purely real $C$ in this way does not have any feed-forward driving.  In this case, measurement results are discarded, and the evolution is just due to measurement backaction (which is equivalent to classical noise).  No entanglement is generated.  Even if $\Im(C)\neq0$, the measurements and feedforward could all be purely local processes.  Again, in this case no entanglement generation is possible, and the environment dissipation would be categorized as being classical.  

{ We also comment on a third way to realize a restricted class of processes with a complex $C$ using classically-stochastic time-dependent system Hamiltonians
~\cite{maassen1987,viola2019}.  Suppose our system is coupled to classical noise whose time integral is a Poisson process with the rate $\nu$.  
Averaging over this noise leads to a term
$\nu (\hat L_k^\dagger \hat\rho \hat L_k -\hat\rho)$ in the master equation, where $\hat{L}_k$ is a unitary operator reflecting the coupling of the noise to the system.  
Combined with the previously discussed classical white-noise processes that generate Hermitian $\hat L_k$, we now have an approach for realizing any master equation with Hermitian or unitary Lindblad operators~\cite{maassen1987,viola2019}. In our case, a process that can be expressed in this way can have non-zero $\Im(C)$, but can never generate system entanglement.  The reverse statement is however, surprisingly, not true:  there exist master equations with non-zero $\Im(C)$ that are not equivalent to the above classical noise model, but nonetheless are unable to generate entanglement (see SM~\cite{supplement}).}  

{\it Entanglement generation-- } We have defined the environment-induced dissipation in  \eqref{eq:lindblad_full} as being quantum if it can generate entanglement within the system.  To study this condition quantitatively, we now employ the Peres-Horodecki (PH) criterion~\cite{peres1996separability,horodecki2001separability}. It ensures that a state whose partial transpose has a negative eigenvalue is entangled. Thus, to see if environmental dissipation can create entanglement, we could in principle evolve arbitrary initial product states, and check whether the PH condition is violated for {\it any} $t>0$. 
We do not require steady-state entanglement generation (unlike e.g.~reservoir engineering protocols~\cite{poyatos1996quantum}).
This would appear to be a formidable task. Fortunately, we can greatly simplify this problem. First, we find the exact evolution of the partial transposed state of the system with respect to a chosen subsystem $A$, i.e.,  $\hat{\rho}^{T_A}=(\mathcal{T}_A\otimes1)(\hat{\rho})$ (see Fig.~\ref{fig:schematic}b). Surprisingly, the evolution $\frac{d\hat{\rho}^{T_A}}{d t}= \tilde{\mathcal{L}}(\hat{\rho}^{T_A})$ is still in the GKSL form $\tilde{\mathcal{L}}(\hat{\rho}^{T_A})= -i[\hat{H}_{\rm{PT}},\hat{\rho}^{T_A}]+\tilde{\mathcal{L}}_{\rm{diss}}(\hat{\rho}^{T_A})$~\cite{supplement}, with $\hat{H}_{\rm{PT}}= \frac{1}{2}\sum_{ij} \tilde{h}_{ij} \hat{Z}_i \hat{Z}_j$, and 
\begin{equation}\label{eq:dissipator_pt}
 \tilde{\mathcal{L}}_{\rm{diss}}(\hat{\rho}^{T_A})=  \sum_{i,j} \tilde{c}_{ij} (\hat{Z}_i \hat{\rho}^{T_A} \hat{Z}_j - \frac{1}{2} \{\hat{Z}_i \hat{Z}_j, \hat{\rho}^{T_A}\}).
\end{equation}
In this partial transposed equation of motion, the coefficients (for $i<j$) of the Hamiltonian and the dissipator are given by
\begin{align}
	\label{eq:pt_hamiltonian}
	&(\tilde{h}_{ij},\tilde{c}_{ij})=\nonumber\\
	&\begin{cases}
		(\Im(c_{ij}),-\Re(c_{ij})+i h_{ij}) & i\in A \text{ and }  j\notin A\\
		(-h_{ij},c_{ji} ) & i\in A \text{ and }  j\in A\\
		(h_{ij},c_{ij}) & \text{otherwise}
	\end{cases},
\end{align}
The coefficients for $i>j$ can be inferred from the symmetries $\tilde{c}_{ij}=\tilde{c}_{ji}^*$ and $\tilde{h}_{ij}=\tilde{h}_{ji}$. 

Note that $\hat{H}_{\rm{PT}}$ is Hermitian, hence the only way that $\tilde{\mathcal{L}}$ can generate non-positive states is through the dissipative part $\tilde{\mathcal{L}}_{\rm diss}$ specified by $\tilde{C}$.  This is in general possible, as $\tilde{C}$ is Hermitian, but not necessarily PSD. If our original master equation had a non-zero Ising Hamiltonian $h_{ij} \neq 0$, we see clearly from Eq.~\eqref{eq:pt_hamiltonian} that $\tilde{C}$ could be non-PSD.  This simply reflects that fact that Hamiltonian Ising interactions can create entanglement.  We are however interested in the effects of the dissipative evolution alone, i.e. $h_{ij}=0$.  As we will see, in this case too $\tilde{C}$ can fail to be positive.   Note there is a small subtlety here:
to show the possibility of entanglement generation, we need to show that the evolution generated by $\tilde{\mathcal{L}}_{\rm diss}$ is not positive.  However, a negative eigenvalue of 
$\tilde{C}$ only implies in general that the evolution is not completely positive (CP)~\cite{wolf2012quantum}.  Luckily in our case we can show that these two notions coincide (see SM \cite{supplement}).  

To summarize, we have found a sufficient condition for an environment to be entangling and hence be truly quantum. To check this condition, we need to find $\tilde{C}$ using the recipe in Eq.~\eqref{eq:pt_hamiltonian} and examine its eigenvalues for all the possible choices of subsystem $A$. If there exist a negative eigenvalue in any of these cases, it indicates that the dissipative evolution in entangling. We note that, however, the absence of a negative eigenvalue does not rule out entanglement generation for $n>2$, as there are entangled states with  a positive partial transpose~\cite{horodecki1998mixed}.

{\it Case studies-- } 
To provide further intuition, we now analyze three special cases of Markovian correlated dephasing on $n$ qubits. Let $\{\mathbf{e}_i\}_{i=0}^{n-1}$ denote the standard basis of $\mathbb{C}^{n}$.  Additionally, define the Fourier basis $\{\mathbf{f}_k\}_{k=0}^{n-1}$, such that $\mathbf{f}_k= \frac{1}{\sqrt{n}}\sum_{j=0}^{n-1} \omega^{jk} \mathbf{e}_j$, where $\omega=\exp(2\pi i/n)$.
As our first example, consider a purely real correlation matrix $C^{(1)}=\mathbf{f}_0 \mathbf{f}_0^\dagger$ whose entries are $c^{(1)}_{ij}=1/n$ for all $i$'s and $j$'s. As mentioned earlier, this process can always be emulated by classically correlated fluctuations of $\sigma_z$ terms in the Hamiltonian and is not entangling. Using our procedure, $\tilde{C}^{(1)}$ stays PSD under any bipartition. This is because for any real $C$, Eq.~\eqref{eq:pt_hamiltonian} is equivalent to mapping $\mathbf{e}_i\to-\mathbf{e}_i$ for all $i$'s in one of the partitions. This is a unitary transformation and does not change the eigenvalues of the originally PSD matrix $C$. 

Next, we consider a correlation matrix $C^{(2)}$, whose 
elements above (below) the diagonal are all $i$ $ (-i)$. The diagonal elements are all set equal to a constant $\gamma=n-1$, chosen so that $C^{(2)}$ is PSD.  This $C$ matrix corresponds to an environment with broken TRS and with vanishing classical noise correlations (a situation where one might expect the bath to be maximally quantum).
For concreteness, we take subsystem $A$ to be the first $m$ of our $n$ qubits. 
Under Eq.~\eqref{eq:pt_hamiltonian}, $\tilde{C}^{(2)}$ is a block diagonal matrix. It is obtained from $C^{(2)}$ by setting the off-diagonal blocks to zero, i.e., $\tilde{c}^{(2)}_{ij}=0$ if $i\in A $ $(i\notin A ) $ and $j\notin A$ $(j\in A)$, and transposing the block corresponding to $A$. Because the original matrix $C^{(2)}$ is PSD and noting that principal submatrices of a PSD matrix are also PSD~\cite{horn2012matrix}, we conclude that $\tilde{C}^{(2)}$ is PSD. We note that the dynamics corresponding to ${C}^{(2)}$ can be fully realized using measurement and feedforward that is local with respect to the $A/B$ bipartition. Hence, despite not being equivalent to classical noise, this environment cannot generate system entanglement, and would be deemed classical under our classification.  

Finally, we analyze a rank-1 complex correlation matrix $C^{(3)}$ that is impossible to emulate with a local measurement and feed-forward strategy. Specifically, we choose $C^{(3)}=\mathbf{f}_1 \mathbf{f}_1^\dagger$.
For $n \geq 3$, we have $\textrm{Im}( C^{(3)}) \neq 0$, corresponding to a bath with broken TRS that mediates non-zero dissipative interactions. 
To show that the corresponding evolution is capable of generating entanglement, it suffices to find one bipartition such that $\tilde{C}^{(3)}$ has a negative eigenvalue. Choosing the first qubit as one of the partitions, results in a rank-3 $\tilde{C}^{(3)}$. As shown in SM~\cite{supplement}, the low-rank nature of $\tilde{C}^{(3)}$ allows us to analytically calculate  { $|\tilde{C}^{(3)}|_+=\frac{2-n}{4n^2}$ (the pseudo-determinant of $\tilde{C}$)}, which implies that $\tilde{C}^{(3)}$ has at least one negative eigenvalue for all $n>2$. Note that in general the
evolution of an initial product state
will at most result in transient entanglement~\cite{plenio2005logarithmic} (but no steady-state entanglement) (see SM~\cite{supplement}).

{\it Random environments-- }
The above examples suggest that both the imaginary and the real parts of the off-diagonals of $C$ are necessary to create entanglement. This expectation is corroborated by examining three-qubit random Liouvillians with $C=w w^\dagger$, where $w$ is drawn from a complex Ginibre ensemble~\cite{ginibre1965statistical} (see SM~\cite{supplement} for details). Analogous ensembles of random Liouvillians have been considered previously, in a different context, to understand spectral properties of random open quantum systems~\cite{sa2020spectral}. Interestingly, the results reported in Ref.~\cite{sa2020spectral} show no sensitivity to the choice of real and complex ensembles, whereas in our work the latter corresponds to broken TRS, which is necessary for entanglement generation. 
\begin{figure}[t]
    \centering
    \includegraphics[width=\columnwidth]{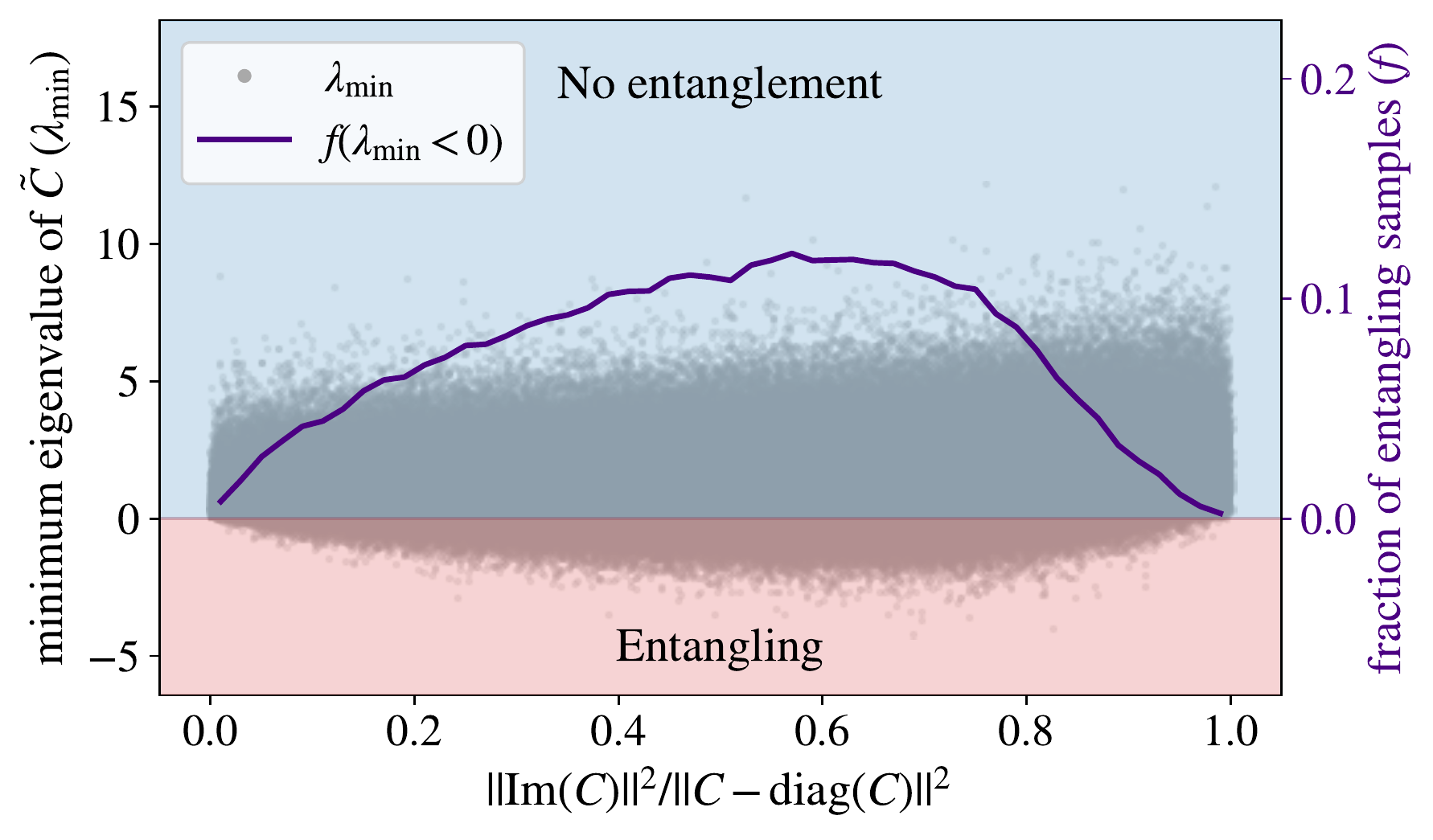}
    \caption{The distribution of the minimum eigenvalue of $\tilde{C}$ for $10^6$ random three-qubit dephasing environments. The solid purple curve (right axis) shows the fraction {(over bins of size $0.02$)} of samples where $\tilde{C}$ has a negative eigenvalue (entangling). A $C$ matrix with purely real or purely imaginary off-diagonals is not entangling. }
    \label{fig:interpolation}
\end{figure}
In Fig.~\ref{fig:interpolation}, we observe that $\lambda_{\rm{min}}$, the minimum eigenvalue of $\tilde{C}$~\footnote{ Note that because of symmetries specifying the partition is not necessary for three qubits}, is non-negative when $C$ is purely real ($\norm{\Im(C)}_{\rm{fro}}=0$), or has a purely imaginary off-diagonal part ($\norm{\Re(C)-{\rm{diag}}(C)}_{\rm{fro}}=0$). However, when the norm of the imaginary and real  off-diagonal parts are comparable, the fraction, $f$, of the samples that are entangling ($\lambda_{\rm{min}}<0$), is maximized.

\begin{figure}[t]
    \centering
    \includegraphics[width=\columnwidth]{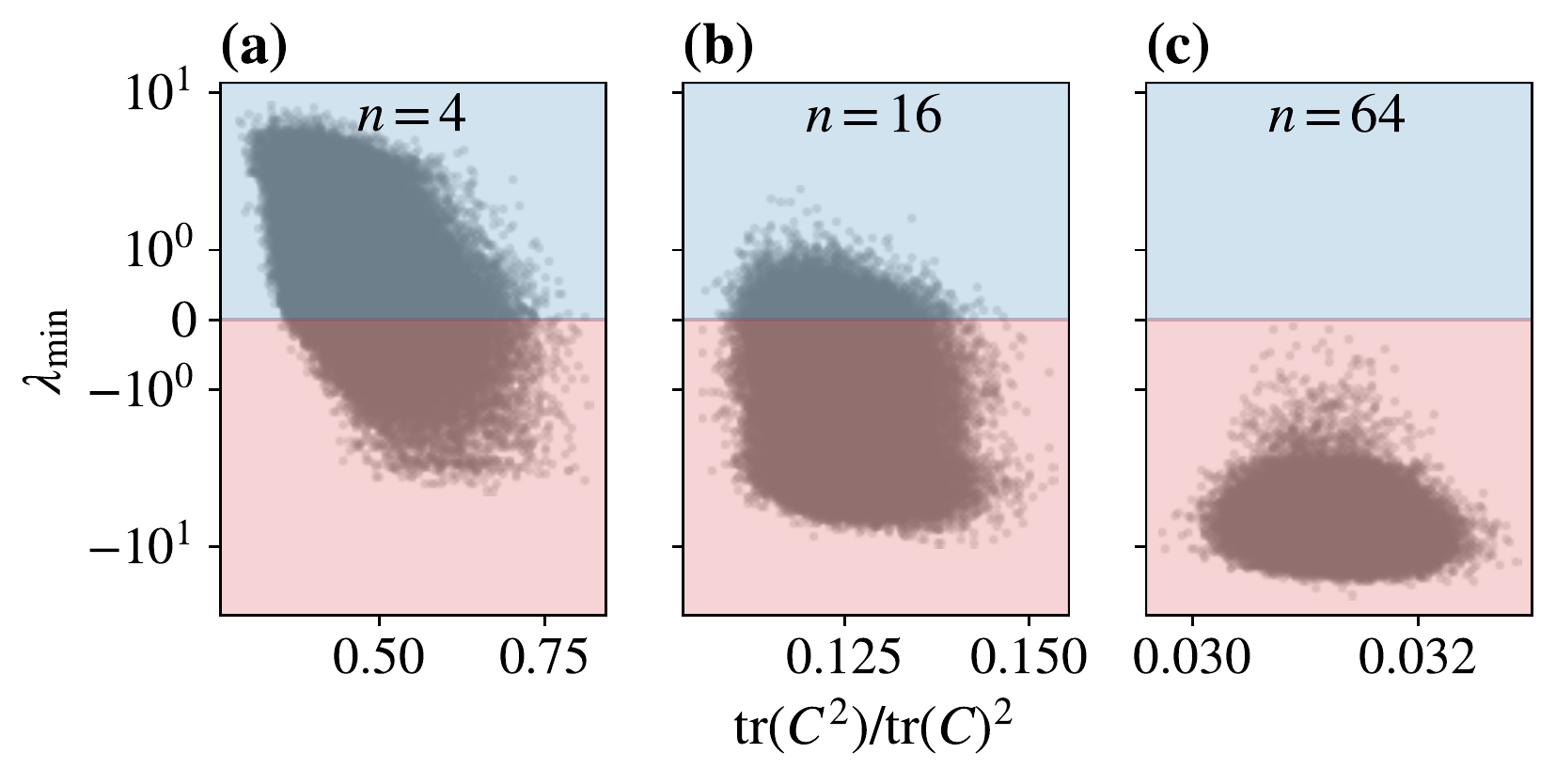}
    \caption{Typicality of entangling $C$s. The distribution of the minimum eigenvalue $\lambda_{\rm{min}}$ of $\tilde{C}$ for $10^5$ random $C$'s as a function of $\tr(C^2)/\tr(C)^2$ for (a) $n=4$, (b) $n=16$, and (c) $n=64$ qubits. To obtain $\tilde{C}$, we choose the first qubit as a subsystem. Entangling $C$'s become more common as the system size increases.}
    \label{fig:typicality}
\end{figure}

Our simple examples also raise another question:  can dissipative interactions only generate entanglement if $C$ is a low rank matrix?  Physically, this corresponds to a situation where the bath couples to the system via only a small number of delocalized system operators.    To answer this question we again consider random $C=ww^\dagger$ and vary system size. We examine the minimum eigenvalue of $\tilde{C}$ acting on the partial-transposed state of the system with respect to the first qubit. To quantify the rank of $C$, we introduce $\tr(C^2)/\tr(C)^2$ that measures variance of $C$ eigenvalues. We observe that the entangling behavior is not a property of rank-1 matrices (for which $\tr(C^2)/\tr(C)^2=1$), and is common in the ensemble of random $C$'s we considered (see Fig.~\ref{fig:typicality}). Note that the minimum eigenvalue of $C$ in this case scales inversely with $n$~\cite{edelman1988eigenvalues}. Hence, the transformation to $\tilde{C}$ is more likely to create a negative eigenvalue (see SM~\cite{supplement}).

{\it Experimental implementation-- } When combined with the ability to measure $C$, our results can serve as a probe for fundamental symmetries and the nature of the environment. The evolution generated by the Lindbladian $\mathcal{L}$ in Eq.~\eqref{eq:lindblad_full} can be decomposed into a decaying part from $\Re(C)$, and a phase evolution from $\Im(C)$ and $\hat{H}_{\rm{LS}}$~\cite{seif2021compressed}. Ref.~\cite{seif2021compressed} presents a compressed sensing protocol that uses randomized measurement to extract both the real and imaginary part of $c_{ij}$ and also the Lamb shift terms $h_{ij}$. Here, we present a less efficient but simpler scheme. Our measurement protocol relies on preparing and measuring Bell states $\ket{\phi_{ij}} = \frac{1}{\sqrt{2}} (\ket{0}_i\ket{0}_j+\ket{1}_i\ket{1}_j)$ on pairs of qubits with $j>i$ (the omitted qubits are assumed to be in the $\ket{0}$ state). Let $\hat{\rho}(t)$ denote the state of the system after some time $t$ that is initially prepared in $\ket{\phi_{ij}}$. The matrix element $\bra{0}_{i}\bra{0}_{j}\hat{\rho}(t)\ket{1}_{i}\ket{1}_{j}$ evolves as $\frac{1}{2}\exp[(i\Omega_{ij}-\Gamma_{ij})t]$. Therefore by measuring this matrix element at different times and finding its decay rate and oscillation frequency we can find both $\Omega_{ij}$ and $\Gamma_{ij}$. The real part of $c_{ij}$ can then be directly extracted from $\Gamma_{ij}=2(c_{ii}+c_{jj}+c_{ij}+c_{ji})$ as shown in  \cite{layden2018spatial}. The analysis of $\Omega_{ij}$, however, is more subtle. First, unlike $\Gamma_{ij}$,  $\Omega_{ij}$ contains linear contributions from $\Im(c_{km})$'s, where $k,m$ are not just restricted to $\{i,j\}$. Therefore, we need to solve a linear system to find $\Im(C)$. Secondly, there might be other sources of phase evolution in addition to $\Im(C)$, e.g., the Lamb shift term,  in the experiment. In particular, it is impossible to distinguish Lamb shift terms from $\Im(C)$ using only the above measurements. However, performing an additional set of measurements using $\ket{\bar{\phi}_{ij}}=(\bigotimes_{k=1}^i \ket{1}_k)\otimes \frac{1}{\sqrt{2}}(\ket{0}_j +\ket{1}_j)$  with $j>i$, and where qubits with omitted index are in $\ket{0}$, provides enough information to distinguish $h_{km}$'s contributions from $\Im(c_{km})$'s~\cite{supplement}.  Determining the nature of such coherent phase errors is also helpful in a broader context.  In the context of error mitigation,  where correlated noise processes severely impact the performance of the device~\cite{palma1996quantum,monz2011coherence,wilen2021correlated,McEwen2021resolving}, it is important to correctly identify the source of noise to combat it. For example, coherent phase errors originating from parasitic $ZZ$ couplings (see e.g.,~\cite{munadada2019suppression,Sundaresan2020reducing}) can be simply canceled by an offset Hamiltonian, whereas a simple offset cannot help with the errors coming from $\Im(C)$.

{\it Discussion-- }Our study reveals that the presence or absence of TRS has a profound effect on dephasing dissipation, even in the Markovian limit:  broken TRS is needed for the dissipation to be entanglement-generating (and hence quantum).
Our work thus provides a concrete experimental protocol for detecting the presence of broken environmental TRS.
Note that to check entanglement generation between all possible bipartitions, our protocol requires a time that scales exponentially with the system size.  It is intriguing to ask whether this is a fundamental limitation, or whether more efficient schemes are possible. It would be extremely interesting to apply our ideas to systems with nonlinear and nonlocal coupling to an environment and going beyond the Markovian limit, and more generally study the role of TRS in more general kinds of dissipative dynamics (e.g.~baths that couple transversely to the qubits). 
\begin{acknowledgments}
{We thank Liang Jiang and Lorenza Viola for helpful discussions and insightful comments on an earlier version of this manuscript.} This work was partially supported by the Army Research Office under Grant 
No.~W911NF-19-1-0380, and by the Center for Novel Pathways to Quantum Coherence in Materials, an Energy Frontier Research Center funded by the Department of Energy, Office of Science, Basic Energy Sciences. A.S. is supported by a Chicago Prize Postdoctoral Fellowship in Theoretical Quantum Science, A.A.C. acknowledges support from the Simons Foundation through a Simons Investigator award 
(Award No.~669487, AC).
\end{acknowledgments}

\bibliographystyle{apsrev4-1}
\bibliography{correlated}

\pagebreak
\appendix

\section{Derivation of the master equation for a classical dephasing environment}

Dissipative evolution can arise as the description of a system driven by classically-fluctuating time-dependent fields. Here we show that in the case of Markovian dephasing such classical fluctuations can only cause exponential decay of coherences and do not lead to any dissipative interactions or phase evolution.  We consider a model where there is no bath, but the qubits are described by a time-dependent Hamiltonian
\begin{equation}
\hat{H}_{\rm{c}} = \sum_i b_i(t) \hat{Z}_i, 
\end{equation}
where $\langle b_i(t) b_j(t')\rangle = \gamma_{ij}\delta(t-t')$ are correlated Gaussian white noise terms, $\Gamma = (\gamma_{ij})$ is the covariance matrix of the noise, and $\hat{Z}_i$ is the Pauli $\hat{\sigma}_z$ operator acting on qubit $i$. To derive a master equation for the noised-averaged evolution it is helpful to first decompose the noise into independent components. Note that because the covariance matrix is symmetric positive semi-definite (PSD), we can find matrices $O$ and $D$ (diagonal) such that $\Gamma = O D O^T$. We can then express the Hamiltonian as 
\begin{equation}
\hat{H}_{\rm{c}} = \sum_i \tilde{b}_i(t) \hat{L}_i, 
\end{equation}
where  $\hat{L}_i = \sum_{j} (D^{1/2} O^T)_{ij} \hat{Z}_j$ are Hermitian operators with the new noise terms such that $\langle \tilde{b}_i(t) \tilde{b}_j(t')\rangle = \delta_{ij}\delta(t-t')$. 

We now average the time evolution over different realizations of noise to eliminate noise variables~\cite{van1992stochastic,gardiner2004quantum}. Specifically, the evolution of the system is now given by a Stratonovich stochastic differential equation 
\begin{equation}
    d\hat{\rho} = -i\sum_k dW_k [\hat{L}_k,\hat{\rho}],
\end{equation}
where $dW_k$'s are independent Wiener increments. Converting this equation to Ito form, and averaging over noise gives the following master equation
\begin{align}
\frac{d \hat{\rho}}{dt} &= -\frac{1}{2}\sum_k [\hat{L}_k,[\hat{L}_k,\hat{\rho}]]\\ &= -\frac{1}{2}\sum_k(\hat{L}_k \hat{L}_k \hat{\rho} - \hat{L}_k \hat{\rho} \hat{L}_k - \hat{L}_k \hat{\rho} \hat{L}_k + \hat{\rho} \hat{L}_k \hat{L}_k) \\ &= \sum_k \hat{L}_k \hat{\rho} \hat{L}_k -\frac{1}{2} \{\hat{L}_k \hat{L}_k,\hat{\rho}\},
\end{align}
which is in the Lindblad form. Going back to the original basis we find
\begin{equation}\label{eq:classical-lindblad}
\frac{d \hat{\rho}}{dt} = \sum_{i,j} c_{ij} (\hat{Z}_i \hat{\rho} \hat{Z}_j - \frac{1}{2} \{\hat{Z}_i \hat{Z}_j, \hat{\rho}\}), 
\end{equation}
where $c_{ij}$ coincides with $\gamma_{ij}$, the covariance of the noise. In this case, these $c_{ij}$ coefficients are real because the covariance matrix of a classical noise process has to be real.  

\section{Measurement and feed-forward}
Here we present a scheme to realize a complex $C$ using measurement and feed-forward control. As shown in Ref.~\cite{metelmann2017nonreciprocal}, it is possible to realize an evolution with Lindblad operators $\hat L=\hat A-i \hat B$ {together with a Hamiltonian $\hat{A}\hat{B}$} by continuously monitoring the Hermitian operator $\hat A$ and applying the Hermitian operator $\hat B$ whose strength depend on the $\hat A$ measurement signal. Specifically, let $I(t)=\sqrt{k}\expval{\hat A(t)}dt+dW(t)$ denote the measurement signal from continuously monitoring the operator $\hat{A}$. This type of monitoring can be obtained by weakly coupling the system to a meter and measuring the state of the meter in $\Delta t$ time intervals, while taking the limit that the coupling strength vanishes as $\Delta t\to 0$. Here, $k$ is the measurement rate and $dW(t)$ is a Wiener increment.  We can then use the measurement record to apply a Hamiltonian $\hat{H}_{FF}(t)=\sqrt{\alpha_{FF}} I(t-\tau) \hat B$, where $\alpha_{FF}$ is a rate determining the strength of the applied force and $\tau$ is the delay.  We are interested in the unconditional evolution, i.e. averaging over different measurement realizations.   In the limit of $\tau\to0^+$ one finds that this evolution is described by~\cite{metelmann2017nonreciprocal}
\begin{equation}
    \frac{d\hat{\rho}}{dt} = \frac{k}{4}\mathcal{D}[\hat A]\hat{\rho} + \alpha_{FF}\mathcal{D}[\hat B]\hat{\rho}-i\frac{\sqrt{k \alpha_{FF}}}{2}[\hat B,\hat A\hat{\rho}+\hat{\rho} \hat A],
\end{equation}
where $\mathcal{D}[\hat L]\hat{\rho}= \hat L\hat{\rho} \hat L^\dagger -\frac{1}{2}\{\hat L^\dagger \hat L,\hat{\rho}\}$. Setting $k=4\alpha_{FF}$ we find that 
\begin{equation}
    \frac{d\hat{\rho}}{dt} =  -i\alpha_{FF}[\hat A \hat B,\hat{\rho}]+\alpha_{FF}\mathcal{D}[\hat A-i\hat B]\hat{\rho}.
\end{equation}
This evolution has a Hamiltonian term in addition to the dissipative evolution that we are interested in. However, by performing the protocol in the reverse direction with a different sign, that is monitoring $\hat B$ and feed-forwarding $-\hat A$ we can cancel the Hamiltonian part and obtain $\frac{d\hat{\rho}}{dt} = \mathcal{D}[\hat A-i\hat B]\hat{\rho}$. 

The above discussion suggests a recipe for engineering a dissipative Linbladian defined by an arbitrary complex $C$ matrix: First, diagonalize $C$ and obtain the Lindblad operators $\hat{L}_k$. For each $\hat{L}_k$ let $\hat{L}_k^+ = \frac{1}{2}(\hat{L}_k+\hat{L}_k^\dagger)$ and $\hat{L}_k^-=-\frac{i}{2}(\hat{L}_k-\hat{L}_k^\dagger)$, corresponding to the Hermitian and anti-Hermitian part, respectively. Letting $\hat A_k=\hat{L}_k^+$ and $\hat B_k= -\hat{L}_k^-$ and performing the above protocol in both forward and reverse directions for each $k$ realizes an evolution equivalent to that generated by $C$. 

\section{Derivation of the partial-transposed evolution}

Here we show how the dynamics of the partial-transposed state are connected to the original evolution. As discussed in the main text, we again focus on multiqubit systems whose dynamics can be described by a master equation of the following form (see also Eq.~\eqref{eq:lindblad_full} in the main text)
\begin{equation}
	\label{seq:lindblad_full.c}
	\mathcal{L}(\hat{\rho})= -i[\hat{H}_{\rm{LS}},\hat{\rho}] + \sum_{i,j} c_{ij} 
	\left(\hat{Z}_i \hat{\rho} \hat{Z}_j - \frac{1}{2} \{\hat{Z}_i \hat{Z}_j, \hat{\rho}\} \right)
	, 
\end{equation}
where $\hat{H}_{\rm{LS}}=\frac{1}{2}\sum_{ij} h_{ij} \hat{Z}_i \hat{Z}_j$ gives the Lamb shift Hamiltonian. We first write the terms in Eq.~\eqref{seq:lindblad_full.c}, taking into account the symmetry constraints on the coefficients ($h _{ij} = h _{ji} $, $c _{ij} = c _{ji}^{*} $), i.e.~we have 
\begin{align}
	\label{seq:lindblad_full.simp}
	\mathcal{L}(\hat{\rho})&= \sum_{i,j}
	-\frac{i}{2} (  h _{ij}- i\Re c_{ij} )
	\hat{Z}_i \hat{Z}_j \hat{\rho}
	\nonumber \\
	&+ \sum_{i,j} \frac{i}{2} (  h _{ij} +i \Re c_{ij} )
	\hat{\rho} \hat{Z}_i \hat{Z}_j  \nonumber \\
	&+ \sum_{i,j} \Re c_{ij} 
	\hat{Z}_i \hat{\rho} \hat{Z}_j 
	\nonumber \\
	& 
	+ i \sum_{i \ne j} \Im c_{ij} 
	\hat{Z}_i \hat{\rho} \hat{Z}_j
	.
\end{align}

\begin{figure}[t]
	\centering
	\includegraphics[width=\columnwidth]{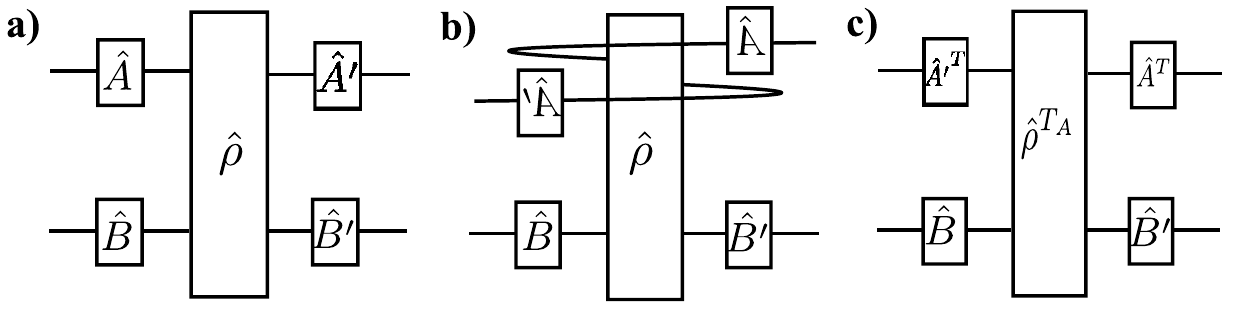}
	\caption{Diagrammatic derivation of the partial-transpose operation. (a) We start with a diagram representing $(\hat{A}  \otimes \hat{B}  ) 
		\hat{\rho}  
		(\hat{A}' \otimes \hat{B}' )$, where the subsystem $A$ corresponds to the upper half of the panel with operators $(\hat{A}$ and $\hat{A}'$. (b) To perform the partial transpose with respect to the subsystem $A$, we transpose only the upper half of the diagram. (c) After rearranging the diagram we find the results to be $(\hat{A} ' {}^{T} 
		\otimes \hat{B}) 
		\hat{\rho} ^{T_A} 
		(\hat{A} ^{T} \otimes \hat{B}' )$. }
	\label{fig:pt_diagram}
\end{figure}

We now consider a generic bipartition that divides the total system into subsystems $A$ and $B$, and we use $\hat{\rho}^{T_A}=(\mathcal{T}_A\otimes1)(\hat{\rho})$ to denote the partial transpose operation with respect to $A$ qubits in the eigenbasis of $\hat Z_i$ ($i \in A$). We note that explicit action of partial transpose is basis-dependent; however, the question of whether a given Markovian master equation can generate entanglement ultimately would not rely on the specific basis used in the partial transpose operation, and here we pick $\hat Z_i$ basis for convenience. We define eigenstates of $\hat Z_\ell$ operators as $\ket{\boldsymbol{\alpha}_{A}, \boldsymbol{\alpha}_{B}}$, where 
$\hat Z_\ell
\ket{\boldsymbol{\alpha}_{A}, \boldsymbol{\alpha}_{B}}
= (\boldsymbol{\alpha} _{\sigma} ) _{\ell}  \ket{\boldsymbol{\alpha}_{A}, \boldsymbol{\alpha}_{B}}$ for $\ell \in \sigma$ ($\sigma = A,B$). Note that contributions to Eq.~\eqref{seq:lindblad_full.simp}, and hence Eq.~\eqref{seq:lindblad_full.c}, can be generally written in the form of 
$ (\hat{A}  \otimes \hat{B}  ) 
\hat{\rho}  
(\hat{A}' \otimes \hat{B}' ) $, where $\hat{A} $, $\hat{A}' $ ($\hat{B} $, $\hat{B} '$) are operators acting on $A$ ($B$). We also introduce a shorthand notation $\boldsymbol{\alpha}$ for the basis vectors of the total system by concatenating the basis for $A$ and $B$ subsystems as  $\boldsymbol{\alpha}^T\equiv [\boldsymbol{\alpha}_{A}^T, 
\boldsymbol{\alpha}_{B}^T]$.
We can then find the partial transpose of $ (\hat{A}  \otimes \hat{B}  ) 
\hat{\rho}  
(\hat{A}' \otimes \hat{B}' )$ as 
\begin{align}
	& (\hat{A}  \otimes \hat{B}  ) 
	\hat{\rho}  
	(\hat{A}' \otimes \hat{B}' )^{T_A} 
	\nonumber \\
	= & \sum _{\boldsymbol{\alpha},
		\boldsymbol{\beta} }
	\rho _{\boldsymbol{\alpha},
		\boldsymbol{\beta} }  (
	\hat{A}  \otimes \hat{B} \ketbra{\boldsymbol{\alpha}_{A}, \boldsymbol{\alpha}_{B}}
	{\boldsymbol{\beta}_{A}, \boldsymbol{\beta}_{B} }
	\hat{A}' \otimes \hat{B}' )^{T_A} \nonumber\\
	= & 
	\sum _{\boldsymbol{\alpha},
		\boldsymbol{\beta} }
	\rho _{\boldsymbol{\alpha},
		\boldsymbol{\beta} } 
	(\hat{A}  
	\ketbra{\boldsymbol{\alpha}_{A}}
	{\boldsymbol{\beta}_{A}}
	\hat{A} ' )^{T} 
	\otimes 
	(\hat{B} 
	\ketbra{ \boldsymbol{\alpha}_{B}}
	{ \boldsymbol{\beta}_{B} }
	\hat{B}' )  \nonumber\\
	= & 
	\sum _{\boldsymbol{\alpha},
		\boldsymbol{\beta} }
	\rho _{\boldsymbol{\alpha},
		\boldsymbol{\beta} } 
	(\hat{A}^{T}   
	\ketbra{\boldsymbol{\beta}_{A}}
	{\boldsymbol{\alpha}_{A}}
	\hat{A} ' {}^{T}  )
	\otimes 
	(\hat{B} 
	\ketbra{ \boldsymbol{\alpha}_{B}}
	{ \boldsymbol{\beta}_{B} }
	\hat{B}' )  \nonumber\\
	= & \sum _{\boldsymbol{\alpha},
		\boldsymbol{\beta} }
	\rho _{\boldsymbol{\alpha},
		\boldsymbol{\beta} } 
	(\hat{A} ' {}^{T} 
	\otimes \hat{B}
	\ketbra{\boldsymbol{\beta}_{A}, \boldsymbol{\alpha}_{B}}
	{\boldsymbol{\alpha}_{A}, \boldsymbol{\beta}_{B} }
	\hat{A} ^{T} \otimes \hat{B}' )  
	\nonumber\\
	= & (\hat{A} ' {}^{T} 
	\otimes \hat{B}) 
	\hat{\rho} ^{T_A} 
	(\hat{A} ^{T} \otimes \hat{B}' )  
	.  
	\label{seq:pt.general}
\end{align}
A diagrammatic derivation of this identity is shown in Fig.~\ref{fig:pt_diagram}.
Making use of the general identity in Eq.~\eqref{seq:pt.general}, it is straightforward to show that terms in Eq.~\eqref{seq:lindblad_full.simp} that only include $B$ qubit operators $ \hat{Z}_i, \hat{Z}_j$ ($i,j \notin A$) are invariant under the partial transpose operation $ {T_A} $. On the other hand, if both $i,j \in A$, the corresponding terms in Eq.~\eqref{seq:lindblad_full.simp} would transform according to simple transpose as $\tilde{h}_{ij} = - h_{ij}$, and $\tilde{c}_{ij}= c_{ji}$. For contributions that include both qubit operators in $A$ and $B$, without loss of generality we may assume $i  \in A$, $j \notin A$, so that from Eq.~\eqref{seq:pt.general} we have 
\begin{align}
	(\hat{Z}_i \hat{Z}_j 
	\hat{\rho})^{T_A} 
	&=   \hat{Z}_j \hat{\rho} ^{T_A}  \hat{Z}_i
	, \\
	( \hat{\rho}
	\hat{Z}_i \hat{Z}_j)^{T_A} 
	&=  \hat{Z}_i \hat{\rho}^{T_A} 
	\hat{Z}_j 
	, \\
	(\hat{Z}_i \hat{\rho} \hat{Z}_j)^{T_A} 
	& =  \hat{\rho}^{T_A} \hat{Z}_i \hat{Z}_j
	, \\
	(\hat{Z}_j \hat{\rho} \hat{Z}_i)^{T_A} 
	& =  \hat{Z}_i \hat{Z}_j \hat{\rho}^{T_A}
	. 
\end{align}
Thus, we obtain $\Im(\tilde{c}_{ij}) =  h_{ij} $, $ \Re(\tilde{c}_{ij}) = - \Re({c}_{ij})$, and $\tilde{h}_{ij} = \Im({c}_{ij}) $. Further making use of the fact that $\Im({c}_{ji}) = -\Im({c}_{ij})$,  $\Re({c}_{ji}) = \Re({c}_{ij})$,  and ${h}_{ji}={h}_{ij}$, we arrive at Eqs.~\eqref{eq:pt_hamiltonian} in the main text.

\section{Equivalence of positivity and complete positivity in multi-qubit dephasing}
We consider the diagonal form of the evolution in Eq.~\eqref{eq:lindblad_diss} in the main text which is given by
\begin{equation}
	\mathcal{L}_{\rm{diss}}(\hat{\rho})=\sum_k \gamma_k (\hat{L}_k \hat{\rho} \hat{L}_k^\dagger -\frac{1}{2} \{\hat{L}_k^\dagger \hat{L}_k,\hat{\rho}\}),
\end{equation}
where all $\hat{L}_k$'s commute with each other as they are linear combination of commuting operators $\hat{Z}_i$. Consider the case where the evolution is not completely positive. Without the loss of generality assume that $\gamma_0 < 0$. For an equation in the general GKSL form, this does not imply that the evolution of the system alone violates positivity. However, we show that in the case of multi-qubit dephasing, complete positivity and positivity are equivalent.  Note that by definition a violation of positivity implies a violation of complete positivity. So, to establish the equivalence, we only need to show that a violation of complete positivity results in a non-positive evolution. 
Consider the infinitesimal evolution 
\begin{equation}
	\hat{\rho}(t+\delta t) = \hat{\rho}(t) + \delta t\mathcal{L}(\hat{\rho}) + \mathcal{O}(\delta t^2),
\end{equation}
which is linear to first order. For a pure initial state at $t=0$ we have 
\begin{widetext}
	\begin{equation}\label{eq:shorttime}
		\hat{\rho}(\delta t) = \ketbra{\psi} + \delta t \sum_k \gamma_k  \left(\hat{L}_k \ketbra{\psi}\hat{L}_k^\dagger -\frac{1}{2}\{\hat{L}_k^\dagger \hat{L}_k, \ketbra{\psi}\}\right)+ \mathcal{O}(\delta t^2).
	\end{equation}
\end{widetext}
Positivity requires that $\expval{\hat{\rho}}{\phi}\geq 0$ for all choices of $\ket{\phi}$. Therefore, we need to find a state that violates this inequality. We can achieve that by choosing 
\begin{align}
	\ket{\psi}&=\ket{+}^{\otimes n},\\
	\ket{\phi} &= \hat{L}_0 \ket{\psi}.
\end{align}
Note that we can normalize the diagonal dissipator $ \hat{L}_0$ such that $\bra{\phi}\phi\rangle  = {\bra{\psi}\hat{L}_0^\dagger \hat{L}_0}\ket{\psi}=1$.  Using the above choice we have 
\begin{align}
	\braket{\phi}{\psi} &= 0, \label{eq:psiphiorth} \\
	\bra{\phi}{\hat{L}_k}\ket{\psi}&= \delta _{k,0}  , 
\end{align}
{since $\hat L_k \ket{\psi}$ is orthogonal to $\ket{\psi}$ as $ \hat{L}_k$ is a linear combination of $\hat{Z}_i$. Moreover, $\bra{\psi} \hat{L}^\dagger_{k'} \hat{L}_k \ket{\psi}$ is non-zero only if $k=k'$ because of the orthogonoality of eigenvectors of the Hermitian matrix $C$ (c.f. Eq.~\ref{eq:lindblad_full} in the main text). Therefore, the $\hat{L}_k \ketbra{\psi}\hat{L}_k^\dagger$ part of Eq.~\eqref{eq:shorttime} only contributes to $\expval{\hat{\rho}}{\phi}$ when $k=0$.
	Next, we show that the terms in the anti-commutator are also zero. This directly follows from Eq.~\eqref{eq:psiphiorth} by noting that for all $k$'s including $k=0$ we have}
\begin{equation}
	\bra{\phi}\hat{L}_k^\dagger \hat{L}_k\ketbra{\psi}\phi\rangle = \langle\phi{\ketbra{\psi}\hat{L}_k^\dagger \hat{L}_k}\ket{\phi} = 0 .
\end{equation}
Therefore we find that 
\begin{align}
	\expval{\hat{\rho}(\delta t)}{\phi} &= \delta t \gamma_0 \expval {\hat{L}_0 \ketbra{\psi} \hat{L}_0^\dagger}{\phi} +\mathcal{O}(\delta t^2) \\
	&=\delta t \gamma_0   + \mathcal{O}(\delta t^2) .
\end{align}
Because $\gamma_0<0$ we can always choose a $\delta t$ such that 
\begin{equation}
	\expval{\hat{\rho}(\delta t)}{\phi} < 0.
\end{equation}
This concludes our proof that positivity and complete positivity are equivalent for multi-qubit dephasing Lindbladians. 


\section{Comparison between different notions of classical versus quantum environments}

There has been immense interest in the question of how one should distinguish between classical versus truly quantum environments~\cite{nori2018,nori2019,franco2019,viola2019,clerk2020,cywinski2020,chen2021}. 
As discussed in the main text, previous works have almost exclusively focused on the question of representation, i.e.~whether it is possible to {\emph{interpret}} the system time evolution under a given environment as generated by purely classical noise. In contrast, our work utilizes an operational definition: we define a genuinely quantum environment as one that generates dissipation with true entangling power on the system. It is important to note that generally, these two approaches to identifying quantum environments are not equivalent to each other. To illustrate this point, here we provide a class of $2$-qubit Markovian dephasing environments that has no entangling power between the two qubits, and yet cannot be realized via stochastic processes generated by purely classical noise.

We now consider dissipation described by the following Lindbladian
\begin{equation}
	\mathcal{L}(\hat{\rho})= 
	\mathcal{D} 
	[ \hat{Z}_1 + r e ^{i\alpha}
	\hat{Z}_2 ] 
	\hat{\rho} 
	, 
	\label{smeq:2qb.gen.rk1}
\end{equation}
where $ r $ and $\alpha $ are real coefficients, and we assume $ 0 \le \alpha < \pi $ without loss of generality. Taking partial transpose of Eq.~\eqref{smeq:2qb.gen.rk1} with respect to the first qubit, we obtain a new Lindbladian with
\begin{align}
	\tilde{h}_{ij} &=
	\begin{cases}
		- r \sin \alpha & i\neq j \\
		0 &  \text{otherwise}
	\end{cases},\\
	\tilde{C} &=  \begin{pmatrix}
		1 & - r \cos \alpha \\
		- r \cos \alpha & r ^{2} 
	\end{pmatrix}.
\end{align} 
It is straightforward to see that the resulting dynamics corresponds to a completely positive trace-preserving (CPTP) map, so that if the initial density matrix
$ \hat{\rho}  \left( 0 \right)$ has no entanglement, the partial transposed density matrix 
$ \hat{\rho} ^{T_{1}} \left( t \right)$ remains nonnegative throughout the time evolution. Because the Peres-Horodecki (PH) criterion~\cite{peres1996separability,horodecki2001separability} is a necessary and sufficient condition for $2$-qubit entanglement, we have proven that the starting nonlocal dissipator 
$  \hat{Z}_1 + r e ^{i\alpha} 
\hat{Z}_2 $ in Eq.~\eqref{smeq:2qb.gen.rk1} cannot generate entanglement.

Interestingly, while the dissipation described by Eq.~\eqref{smeq:2qb.gen.rk1} can never give rise to system entanglement, it also could not be simulated using stochastic Hamiltonian dynamics generated by completely classical noise. To show this, we borrow the concept of ``essentially classical (commutative) quantum Markov semigroup" in Refs.~\cite{maassen1987,viola2019}, which encompasses dynamics generated by any Lindbladian $ \mathcal{L} $ that allows following decomposition~\cite{maassen1987}
\begin{equation}
	\mathcal{L}_{\mathrm{e.c.}}(\hat{\rho})
	= -i[\hat{H} ,\hat{\rho}] 
	+ \sum_{l} \mathcal{D} 
	[\hat{M} _{l} ] \hat{\rho} 
	+ \sum_{j} \kappa_{j}  \mathcal{D} 
	[\hat{U} _{j} ] \hat{\rho} 
	. 
	\label{smeq:es.cl.gen}
\end{equation}
Here $  \mathcal{D} [\hat{A}  ]  \hat{\rho} \equiv 
(\hat{A}  \hat{\rho} \hat{A} ^\dagger 
- \{\hat{A} ^\dagger \hat{A}
, \hat{\rho}\} /2  ) $ denotes the standard Lindblad dissipator, $ \hat{H}$ and $ \hat{M} _{l} $ are generic Hermitian operators, and $ \hat{U} _{j} $ are unitary operators. We now show that the dissipator in Eq.~\eqref{smeq:2qb.gen.rk1} with a nonzero $ r $ cannot be rewritten in the above form, unless $ \alpha =0 $ or $\pi/2$. For this specific case, we only need to consider purely dissipative decompositions, with jump operators of the form 
\begin{align}
	\hat{M} _{l} 
	&= m _{l} \hat{Z}_1 
	+  n _{l}  
	\hat{Z}_2 
	, \\
	&= u _{j} \hat{Z}_1 
	+ i v _{j}  
	\hat{Z}_2
	,  
\end{align}
where $ m _{l}$, $ n _{l}$, $ u _{j} $ and $ v _{j} $ are real coefficients. 
Letting the Lindbladian in Eq.~\eqref{smeq:2qb.gen.rk1} to be equal to Eq.~\eqref{smeq:es.cl.gen}, the coefficients must satisfy following conditions 
\begin{align}
	& \sum_{l} m _{l} ^2
	+ \sum_{j} u _{j} ^2
	= 1   
	\label{smeq:decomp.cond.11}
	, \\
	& \sum_{l} n _{l} ^2
	+ \sum_{j} v _{j} ^2
	= r ^{2} 
	\label{smeq:decomp.cond.22}
	, \\
	& \sum_{l} m _{l} n _{l} 
	+ i \sum_{j}  u _{j} v _{j} 
	= r e ^{i\alpha} 
	. 
	\label{smeq:decomp.cond.cr}
\end{align}
Making use of Eq.~\eqref{smeq:decomp.cond.cr} and the Cauchy-Schwarz inequality, we further have 
\begin{align}
	&r ^{2} 
	=\left (\sum_{l} m _{l} n _{l} \right)^2 
	+ \left (\sum_{j}  u _{j} v _{j} \right)^2 
	, \\
	\Rightarrow
	& r ^{2} 
	\le  \left (\sum_{l} m _{l}^2 \right)
	\left (\sum_{l} n _{l}^2   \right)
	+ \left (\sum_{j}  u _{j} ^2 \right)
	\left (\sum_{j}  v _{j}^2  \right)
	.  
\end{align}
Combine the inequality above with Eqs.~\eqref{smeq:decomp.cond.11} and \eqref{smeq:decomp.cond.22}, we obtain
\begin{align}
	& \left (\sum_{l} m _{l}^2 
	+\sum_{j}  u _{j} ^2  \right)
	\left (\sum_{l} n _{l}^2 
	+\sum_{j}  v _{j} ^2  \right)
	\\
	\le  &   
	\left (\sum_{l} m _{l}^2 \right)
	\left (\sum_{l} n _{l}^2   \right)
	+ \left (\sum_{j}  u _{j} ^2 \right)
	\left (\sum_{j}  v _{j}^2  \right)
	,   
\end{align}
which only holds if following relations are valid
\begin{align}
	& \left (\sum_{l} m _{l}^2  \right)
	\left ( \sum_{j}  v _{j} ^2  \right)
	= \left ( \sum_{l}  n _{l} ^2  \right)
	\left ( \sum_{j}  u _{j} ^2  \right)
	= 0
	.   
\end{align}
However, the equation above requires either $ m _{l} \equiv 0$ or $ v _{j} \equiv 0 $, which would contradict with Eq.~\eqref{smeq:decomp.cond.cr} if both $  \cos \alpha$ and 
$  \sin \alpha  $ are nonzero (i.e.~$ \alpha$ is not equal to $0 $ or $\pi/2$). As such, we have shown that the nonlocal dissipator $ \hat{Z}_1 + r e ^{i\alpha} 
\hat{Z}_2 $ ($\alpha \ne 0, \pi/2$) cannot be simulated by random unitary dynamics associated with classical noise models.  

Going beyond this two-qubit example, we remark that a correlated dephasing process that can be expressed in terms of a combination of diffusive and Poisson processes cannot generate entanglement.  For such systems, our notion of classicality coincides with that of Ref.~\cite{viola2019}. Moreover, we do not rule out entanglement generation by non-local Poisson processes that are not in the form of correlated dephasing considered here. We leave a thorough investigation into how the two notions of quantum environments differ in more general scenarios to future works.


\section{Violation of complete positivity by rank-1 Lindbladians}
Assume that we have a single Lindblad operator $\hat{L}=\frac{1}{\sqrt{n}}\sum_{k=0}^{n-1} \omega^k \hat{Z}_k$ acting on $n$ qubits, where $\omega=\exp(2\pi i/n)$. The matrix $C$ (c.f. Eq.~\eqref{eq:lindblad_full} in the main text) for this process is given by $c_{jk} =\frac{1}{{n}} \omega^{j-k}$. Let $\mathbf{f}_k=\frac{1}{\sqrt{n}}\sum_j \omega^{jk} \mathbf{e}_j$, where $(\mathbf{e}_j)_i = \delta_{i,j}$ is the standard basis. Note that $\mathbf{f}_k^\dagger \mathbf{f}_{k'}= \delta_{k,k'}$.  Using this notation we can see that $C=\mathbf{f}_1 \mathbf{f}_1^\dagger$. 

We identify qubit $j=0$ as subsystem $A$, and the remaining qubits as subsystem $B$.  
Performing the partial transpose with respect to the first qubit results in a Lindbladian with \begin{equation}
\tilde{c}_{jk}= \begin{cases}
-\Re{c_{jk}} & j=0\text{ or } k=0 \text{ but not both} \\
c_{jk} & \text{otherwise }  
\end{cases}.
\end{equation}
In the following, we want to show that $\tilde{C}$ has a negative eigenvalue. 

We can express $\tilde{C}  = C + \frac{1}{n}(A_r + A_i)$, where
\begin{equation}
A_i =i 
\begin{pmatrix}
0 & \vline &
  \begin{matrix}
  \Im{\omega} & \dots & \Im{\omega^{n-1}}
  \end{matrix} \\
\hline
  \begin{matrix}
  -\Im{\omega}  \\
  \vdots  \\
  -\Im{\omega^{n-1}} 
  \end{matrix} & \vline & 0
\end{pmatrix},
\end{equation}
and \begin{equation}
A_r = -2 
\begin{pmatrix}
0 & \vline &
  \begin{matrix}
  \Re{\omega} & \dots & \Re{\omega^{n-1}}
  \end{matrix} \\
\hline
  \begin{matrix}
  \Re{\omega}  \\
  \vdots  \\
  \Re{\omega^{n-1}} 
  \end{matrix} & \vline & 0
\end{pmatrix},
\end{equation}
are both Hermitian matrices. We can then see that ${\rm{col}}(A_i)={\rm{span}}(\{{\mathbf{e}_0},{\mathbf{s}_1}\})$, where ${\mathbf{s}_1}=\sum_j \Im(\omega^{j}) {\mathbf{e}_j}$. Therefore, ${\rm{rank}}(A_i)=2$. Let ${\bar{\mathbf{s}}_1}=\frac{1}{\sqrt{n/2}}{\mathbf{s}_1}$. We can construct an orthonormal basis $\{{\mathbf{e}_0},{\bar{\mathbf{s}}_1}\}$, in which $A_i = -\sqrt{\frac{n}{2}} \sigma_y$, where $\sigma_y$ is a Pauli matrix. 

Similarly, ${\rm{col}}(A_r)={\rm{span}}(\{{\mathbf{e}_0},{\mathbf{c}_1}\})$, where ${\mathbf{c}_1}=\sum_j \Re(\omega^{j}) {\mathbf{e}_j}$. Therefore, $A_r$ is also a rank-2 matrix. Let ${\bar{\mathbf{c}}_1} = \sqrt{\frac{2}{n-2}} ({\mathbf{c}_1}-{\mathbf{e}_0})$. Note that $\mathbf{e}_0^\dagger \bar{\mathbf{c}}_1=0$. Therefore, $\{{\mathbf{e}_0},{\bar{\mathbf{c}}_1}\}$ forms an orthonormal basis, in which $A_r=-2\sqrt{\frac{n-2}{2}} \sigma_x$, where $\sigma_x$ is a Pauli matrix.

Finally, note that $\{{\mathbf{e}_0},{\bar{\mathbf{s}}_1},{\bar{\mathbf{c}}_1}\}$ forms an orthonormal basis. In this basis, ${\mathbf{f}_1}=\frac{1}{\sqrt{n}}({\mathbf{e}_0}+i\sqrt{\frac{n}{2}} {\bar{\mathbf{s}}_1}+\sqrt{\frac{n-2}{2}}{\bar{\mathbf{c}}_1})$. Since, $A_r$ and $A_i$ already have a simple representation in this basis, we find that $\tilde{C} = \mathbf{f}_1\mathbf{f}_1^\dagger+ \frac{1}{n}(A_r + A_i)$ also takes a simple form 
\begin{equation}
    \tilde{C} = \frac{1}{n}\begin{pmatrix}
    1 & 0 & -\sqrt{\frac{n}{2}-1} \\
    0 & \frac{n}{2} & i\frac{\sqrt{n(n-2)}}{2} \\
    -\sqrt{\frac{n}{2}-1} & -i\frac{\sqrt{n(n-2)}}{2}  & \frac{n}{2}-1
    \end{pmatrix}.
\end{equation}
{We can see that the product of eigenvalues of this matrix (the pseudo-determinant of $\tilde{C}$) is  $|\tilde{C}|_{+} =\frac{2-n}{4n^2}$, which is negative for any $n>2$. }This proves that $\tilde{C}$ has at least one negative eigenvalue. 

\begin{figure}[t]
    \centering
    \includegraphics[width=\columnwidth]{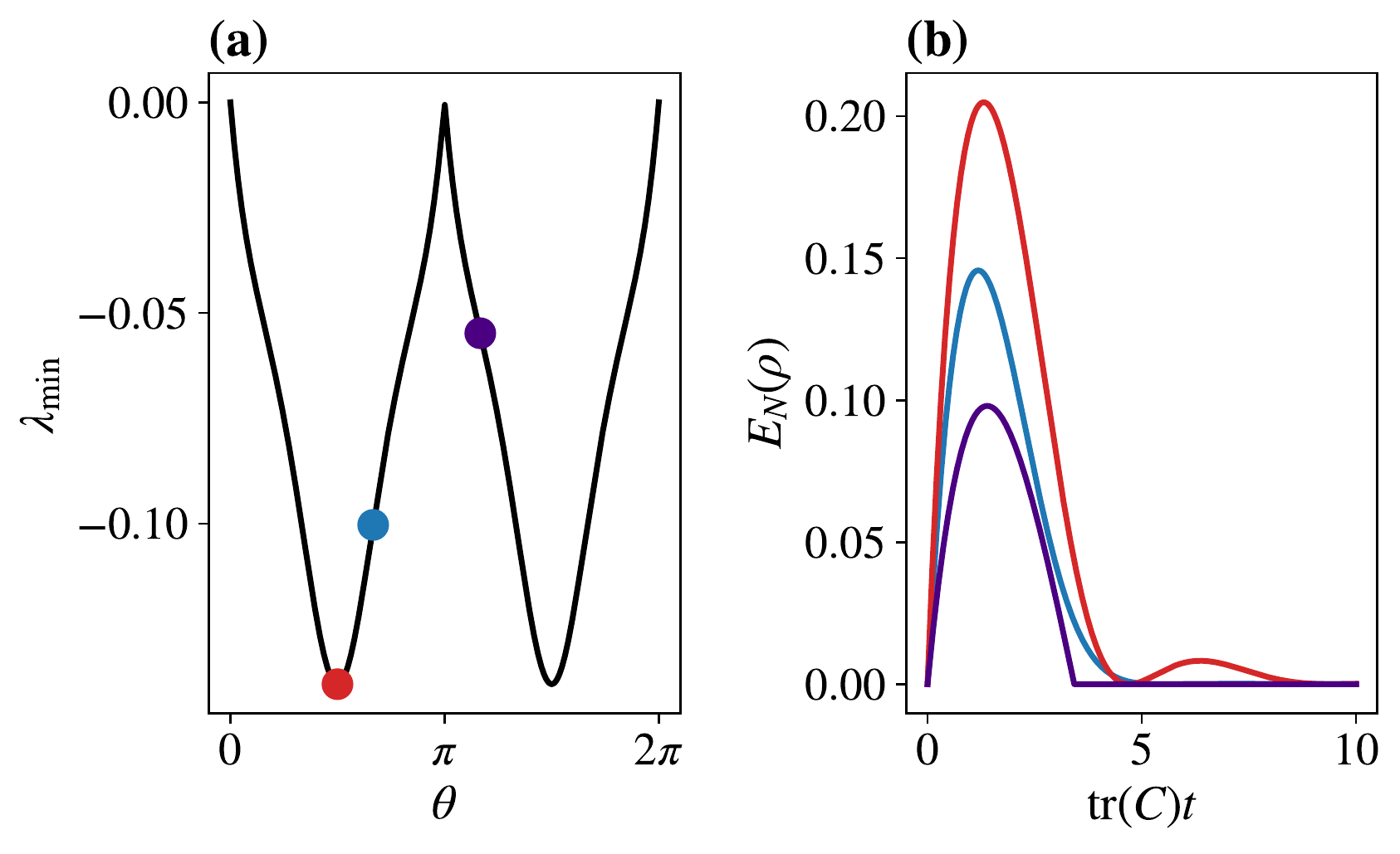}
    \caption{Entangling properties $C=\mathbf{g}(\theta)\mathbf{g}(\theta)^\dagger$ with $\mathbf{g}(\theta)$ defined in Eq.~\eqref{eq:gtheta}. (a) The minimum eigenvalue of $\tilde{C}$ ($\lambda_{\rm{min}}$) as a function of $\theta$ shows that such a dissipator is entangling everywhere except at $\theta=0,\pi,2\pi$. (b) The dynamics of logarithmic negativity ($E_N$) for the values of $\theta$ corresponding to circles in panel (a) shows that the entanglement is transitionary and is not a property of the steady-state. In fact, only the process corresponding to the blue point/curve with $\theta=2\pi/3$ has a dark state.}
    \label{fig:neg_t}
\end{figure}

We now examine a generalized version of this example in more details for the specific case of three qubits. Consider the evolution under a master equation (Eq.~\eqref{eq:lindblad_full} in the main text) with $C=\mathbf{g}(\theta) \mathbf{g}(\theta)^\dagger$, where
\begin{equation}\label{eq:gtheta}
    \mathbf{g}(\theta)=\frac{1}{\sqrt{3}}\begin{bmatrix}
1 & e^{i\theta} & e^{2 i \theta}
\end{bmatrix}^T.
\end{equation}
In Fig.~\ref{fig:neg_t}a we show the minimum eigenvalue of $\tilde{C}$ (with respect to the first qubit) as a function of $\theta$. We observe that all such rank-1 dissipators are entangling except for $\theta=0,\pi,2\pi$, in which $C$ is real. Moreover, we examine the evolution of logarithmic negativity ($E_N$)~\cite{plenio2005logarithmic} of a system with respect to initially prepared in $\ket{+++}$ with respect to the first qubit. As it can be seen in Fig.~\ref{fig:neg_t}b the entanglement appears early in the evolution and vanishes in the steady-state as $t\to\infty$. This signifies the difference between our protocol and dissipation engineering. In fact, except the blue point/curve, with $\mathbf{g}(\frac{2\pi}{3})=\mathbf{f}_1$, the chosen points do not have a dark state. 

\section{Random environments}
In this section we present details of the numerical simulation of random environments in the main text and further investigate their entangling property. 

We study random Liouvillians in the form of Eq.~\eqref{eq:lindblad_full} in the main text with $H_{\rm{LS}}=0$ and $C=w w^\dagger$, where $w$ is an $n\times n$ matrix taken from the complex Ginibre ensemble, i.e., $P(w)\propto\exp[-\frac{1}{2} \tr(w w^\dagger)]$\cite{sa2020spectral}.  In other words, to sample a $w$, we choose the real and imaginary part of each element independently at random from a Gaussian distribution with zero mean and unit variance $N(0,1)$. 

To generate Fig.~\ref{fig:interpolation} in the main text, we generate $10^6$ samples of such random matrices for three qubits, find $\tilde{C}$~\eqref{eq:pt_hamiltonian} that determines the evolution of the partial transposed state with respect to the first qubit, and calculate its minimum eigenvalue. We then plot this minimum eigenvalue for the samples as a function of the relative norm of the imaginary part to the norm of the off-diagonal part, i.e., $||\Im(C) ||_{\rm{fro}}/||C -{\rm{diag}}(C)||_{\rm{fro}}$. Note that for a purely real $C$ this quantity is 0 and for a $C$ with purely imaginary off-diagonal elements (the diagonal elements have to be real) this quantity is 1.  We also show the fraction of the samples that have a negative eigenvalue $f(\lambda_{\rm{min}}<0)$ in bins of size 0.02 (the purple curve in Fig.~\ref{fig:interpolation}).
\begin{figure}[t]
	\centering
	\includegraphics[width=\columnwidth]{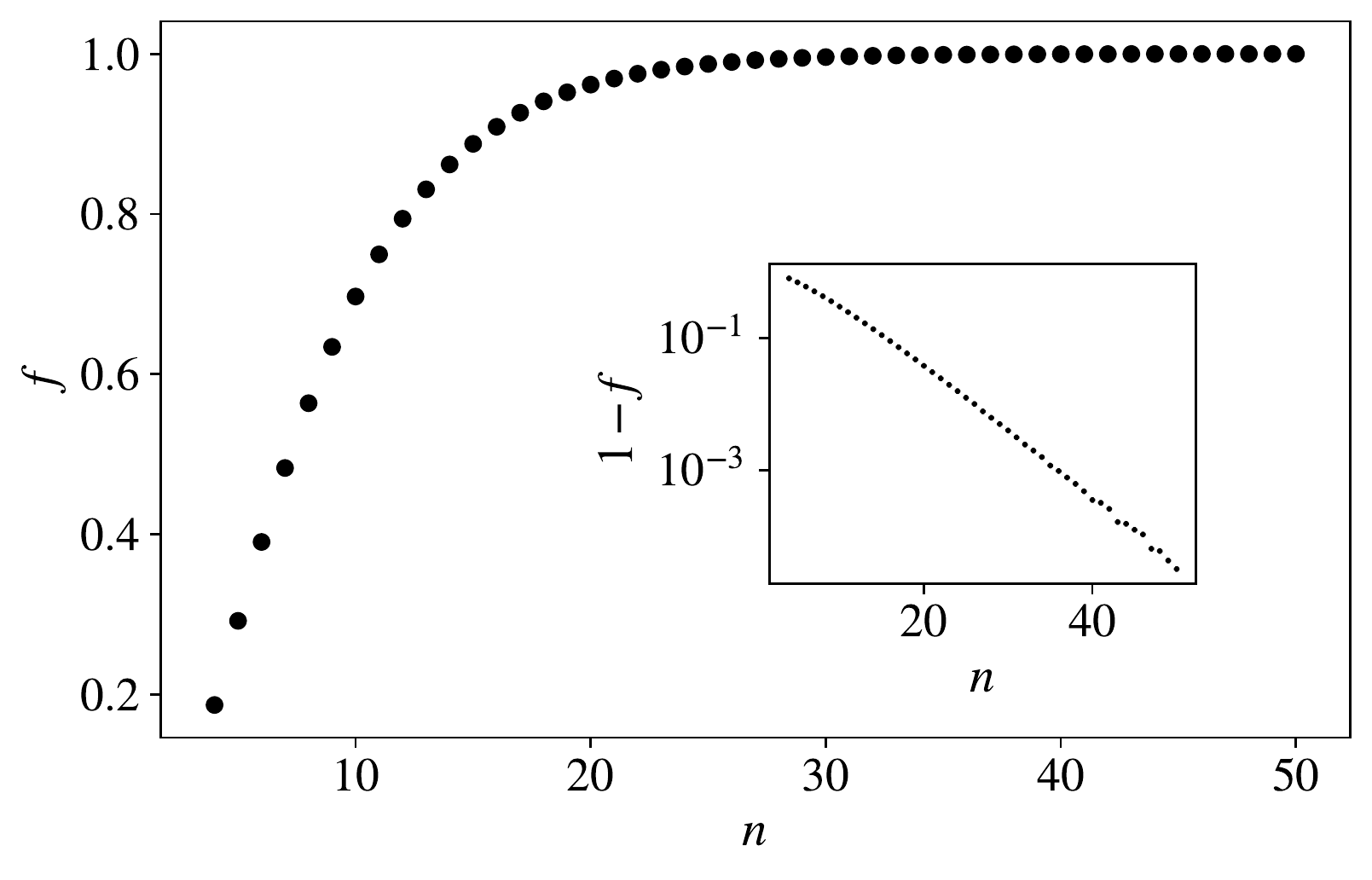}
	\caption{The fraction of random $C=ww^\dagger$ that has a negative eigenvalue, $f$, as a function the system size $n$. This fraction converges exponentially to 1 with system size as shown in the inset. Each point is obtained from $10^6$ samples. }
	\label{fig:convergence_f}
\end{figure}
Fig.~\ref{fig:typicality} is obtained by generating $10^5$ samples of $w$ for $n=4,16,64$ using the same procedure described above. After calculating the minimum eigenvalue of $\tilde{C}$ with respect to the first qubit, we plot the result as a function of  $\tr(C^2)/\tr(C)^2$. To further investigate the entanglement generation in this class of random processes we generate $10^6$ samples for $n=4,5,\dots,50$ and calculate $f(\lambda_{\rm{min}}<0)$ for each sample and plot the results in Fig.~\ref{fig:convergence_f}. We observe that the $f$ converges exponentially to 1. Understanding this interesting behavior is beyond the scope of current manuscript and is left for future work.

\section{Details of the experimental implementation}
\begin{figure}[t]
	\centering
	\includegraphics[width=\columnwidth]{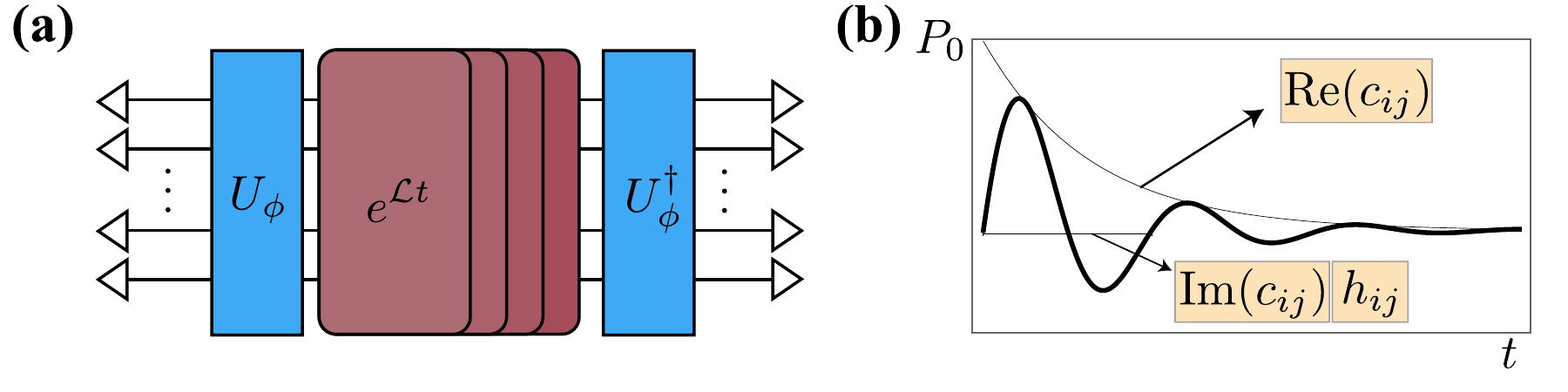}
	\caption{The experimental implementation of the measurements. (a) We prepare the state $\ket{\phi}$ by applying $\hat{U}_\phi$, let it evolve under $\mathcal{L}$ and measure its overlap with the initial state. (b) The overlap is a damped oscillatory function whose frequency is related to $\Im(C)$ and $\hat{H}_{\rm{LS}}$ and its damping rate is related to $\Re(C)$.}
	\label{fig:measurement}
\end{figure}

The evolution of the system under $\mathcal{L}$ (Eq.~\eqref{eq:lindblad_full} in the main text) in the eigenbais of $\hat{Z}_i$'s, i.e.,  $\hat{Z}_i \ket{\boldsymbol{\alpha}} = \alpha_i\ket{\boldsymbol{\alpha}} $ with $\alpha_i=\pm1$, takes a simple form. In particular, $\mathcal{L}(\ketbra{\boldsymbol{\alpha}}{\boldsymbol{\beta}}) = (i\Omega_{\boldsymbol{\alpha\beta}}-\Gamma_{\boldsymbol{\alpha\beta}})\ketbra{\boldsymbol{\alpha}}{\boldsymbol{\beta}}$, where~\cite{seif2021compressed}
\begin{align}
    \Gamma_{\boldsymbol{\alpha\beta}} &= \sum_{km} (\alpha_k - \beta_k) \frac{\Re(c_{km})}{2}(\alpha_m - \beta_m), \label{eq:decayrate} \\
    \Omega_{\boldsymbol{\alpha\beta}} &=\sum_{k<m}(\alpha_k \beta_m-\alpha_m \beta_k) \Im(c_{km}) \label{eq:freqslind}\\ \nonumber
    &- \sum_{k<m} (\alpha_k \alpha_m-\beta_k \beta_m) h_{km}.
\end{align}
Using this notation, $\ket{\phi_{ij}}$ with $i<j$ as defined in the main text corresponds to a superposition of $\frac{1}{\sqrt{2}}(\ket{\boldsymbol{\alpha}}+\ket{\boldsymbol{\beta}})$ with  $\alpha_k = 1$ for all $k$'s and $\beta_k = -1$ for $k=i,j$ and $\beta_k = 1$ otherwise.  Similarly, $\ket{\bar{\phi}_{ij}}$  with $i<j$ corresponds to $\frac{1}{\sqrt{2}}(\ket{\boldsymbol{\alpha}}+\ket{\boldsymbol{\beta}})$ with  $\alpha_k = \beta_k = -1$ for $k\leq i$, $\alpha_k=\beta_k=1$ for $k> i$ and $k\neq j$, while $\alpha_k =-\beta_k=1$ for $k=j$.  Because in the measurement protocol discussed in the main text, we are only concerned with these  classes of states that are fully specified by $i$ and $j$, without introducing an ambiguity, we can use $\Gamma_{ij}$ and $\bar{\Gamma}_{ij}$ to denote the decay associated with $\ket{\phi_{ij}}$ and $\ket{\bar{\phi}_{ij}}$, respectively. Similarly, we use  $\Omega_{ij}$ and $\bar{\Omega}_{ij}$ to denote the phase evolution associated with $\ket{\phi_{ij}}$ and $\ket{\bar{\phi}_{ij}}$, respectively, see Fig.~\ref{fig:measurement}.

Using Eq.~\eqref{eq:decayrate}, we can see that
$\Gamma_{ij}=2(c_{ii}+c_{jj}+c_{ij}+c_{ji})$. Therefore, preparing $\ket{\phi_{ij}}$ and measuring its decay rate under $\mathcal{L}$ (Eq.~\eqref{eq:lindblad_full} in the main text) for all pairs of qubits, together with the information about the single qubit dephasing rates $c_{ii}$ completely determines $\Re(C)$. 

We simplify the expression for $\Omega_{ij}$ by defining $\mathbf{h}$ and $\mathbf{t}$ as the vectorized upper-triangular part of $h_{km}$ and $\Im(c_{km})$,  ordered in increasing $k$ and $m$ such that $k<m$, that is $\mathbf{h} = \begin{bmatrix}
h_{12} & h_{13}& \dots & h_{n-1,n}
\end{bmatrix}^T$ and $\mathbf{t} = \begin{bmatrix}
\Im(c_{12}) & \Im(c_{13}) & \dots & \Im(c_{n-1,n})
\end{bmatrix}^T$. Therefore, we can express $\Omega_{ij} = -\mathbf{h}^T\mathbf{q}^{(ij)} +\mathbf{t}^T\mathbf{w}^{(ij)}$. Here,  $\mathbf{q}^{(ij)}$ and $\mathbf{w}^{(ij)}$ are measurement vectors whose elements are given by
\begin{equation}
	(q^{(ij)}_{km},w^{(ij)}_{km})=\begin{cases}
		(0,0) & k\in  \{i,j\} \text{ and } m\in \{i,j\} \\
		(0,0) & k\notin  \{i,j\} \text{ and } m\notin \{i,j\} \\
		(2,2) & k\in  \{i,j\} \text{ and } m\notin \{i,j\} \\
		(2,-2) & k\notin  \{i,j\} \text{ and } m\in \{i,j\} \\
	\end{cases}.
\end{equation}
Similarly, we have $\bar{\Omega}_{ij} = -\mathbf{h}^T\bar{\mathbf{q}}^{(ij)} +\mathbf{t}^T\bar{\mathbf{w}}^{(ij)}$ where 
\begin{equation}
	(\bar{q}^{(ij)}_{km},	\bar{w}^{(ij)}_{km})= \begin{cases}
		(-2,2) & k\leq i \text{ and } m=j \\
		(2,-2) & i<k<j \text{ and } m=j \\
		(2,2) & k= j \text{ and } m> j\\
		(0,0) & \text{otherwise}
	\end{cases}.
\end{equation}
Therefore, by using $n(n-1)$  distinct measurements of $\Omega_{ij}$ and $\bar{\Omega}_{ij}$ on $n$ qubits for $i<j$ and solving a linear system of equations we can find all of the $\Im(c_{km})$'s and $h_{km}$'s.

\end{document}